\documentclass[runningheads,a4paper]{llncs}

\usepackage[british]{babel}
\usepackage[style=apa, natbib=true, backend=biber]{biblatex}
\DefineBibliographyStrings{british}{%
  andothers = {\emph{et al.},},
}
\renewcommand{\cite}[1]{\citep{#1}}

\DeclareLanguageMapping{british}{british-apa}

\usepackage[ruled]{algorithm2e}
\usepackage{aliascnt}
\usepackage{amsmath,amssymb,mathtools}
\usepackage{rotating}
\usepackage{graphicx}
\usepackage{xcolor}
\usepackage{colortbl}
\usepackage{array}
\usepackage{booktabs}
\usepackage{lscape}
\usepackage{hyperref}
\definecolor{darkblue}{rgb}{0,0,0.4}
\hypersetup{
  colorlinks=true,
  citecolor=darkblue,
  linkcolor=darkblue,
  urlcolor=darkblue
  }
\usepackage{doi}
\usepackage{caption}

\SetAlFnt{\small}
\SetAlCapFnt{\small}
\SetAlCapNameFnt{\small}
\SetAlCapHSkip{0pt}
\IncMargin{-\parindent}

\newcommand{\keywords}[1]{\par\addvspace\baselineskip
\noindent\keywordname\enspace\ignorespaces#1}

\begin{document}


\title{Imposing Higher-Level Structure in Polyphonic Music Generation using Convolutional Restricted Boltzmann Machines and Constraints} 

\titlerunning{Imposing Structure in Music Generation}

%
%
\author{Stefan Lattner$^{1,2}$, Maarten Grachten$^{1}$ \and Gerhard Widmer$^{1,2}$}
%
\authorrunning{Stefan Lattner, Maarten Grachten, Gerhard Widmer}

\institute{$^1$Department of Computational Perception, Johannes Kepler University, Linz, Austria \\
$^2$Austrian Research Institute for Artificial Intelligence, Vienna, Austria}
  

\maketitle

\begin{abstract}
We introduce a method for imposing higher-level structure on generated, polyphonic music.
A Convolutional Restricted Boltzmann Machine (C-RBM) as a generative model is combined with gradient descent constraint optimisation to provide further control over the generation process.
Among other things, this allows for the use of a ``template'' piece, from which some structural properties can be extracted, and transferred as constraints to the newly generated material.
The sampling process is guided with Simulated Annealing to avoid local optima, and to find solutions that both satisfy the constraints, and are relatively stable with respect to the C-RBM.
Results show that with this approach it is possible to control the higher-level self-similarity structure, the meter, and the tonal properties of the resulting musical piece, while preserving its local musical coherence.
\end{abstract}

%
%






\section{Introduction}\label{sec:introduction}
For centuries, mathematical formalisms have been used to generate musical material \cite{Kirchmeyer:1968vk}.
Since computers can automate such processes, automatic music generation has become a small, but steadily emerging field in Artificial Intelligence and Machine Learning.
Nevertheless, automatic music generation as a problem is far from solved: musical outputs created by artificial systems are regarded as a curiosity by human listeners at best, but all too often they are taken as a direct offence to our sense of musical aesthetics.
This sensitivity to violations of even the most subtle musical norms illustrates how complex the problem of (especially polyphonic) music generation is.
In addition, there are only a few objective evaluation criteria to rigorously test and compare music generation systems, all of which involve human judgement \cite{jordanous2012standardised,pearce2001towards}.



This is lamentable, not least since successful methods for automatic music generation would be of considerable commercial interest to the music, gaming and film industries.
Moreover, potential applications that have remained unexplored as of yet, including adaptive music in cars or in fitness applications, could personalise music and thus provide a completely new listening experience.

In line with a global surge in deep learning and neural network modelling over the past decade, several studies address the task of music modelling as a form of sequence learning, in which musical pieces are formulated as a time series of musical events, using state-of-the-art sequence models such as Recurrent Neural Networks (RNN) and Long Short-Term Memory (LSTM).
For restricted genres or representations such as monophonic folk melodies~\cite{sturm16:_music_trans_model_compos_using_deep_learn}, symbolic chord sequences, or drum tracks~\cite{choi16:_text_lstm_autom_music_compos} and even in polyphonic music with clearly defined melodic voices, such as Bach chorales~\cite{boulanger2012modeling,hadjeres2017deepbach, liang2017automatic, huang2017counterpoint},
 sequence modelling approaches yield impressive results that are sometimes hard to distinguish from human-composed material.

However, in more complex musical material, such as piano music from the classical (e.g. Mozart) or romantic period (e.g. Chopin, Liszt), not to mention orchestral works, important musical characteristics may defy straight-forward time series modelling approaches.
\emph{Tonality} for example, is the characteristic that music is perceived to be in a particular (possibly time-variant) \emph{musical key}, implying that some pitches are regarded as more stable than others.
Although the perception of musical key is a complex topic in itself, there is evidence that an important determining factor is the frequency of occurrence of pitches in the piece~\cite{Smith:2004in}.

In addition to tonality, \emph{meter} is an important aspect of music.
Perception of meter is the sensation that musical time can be divided into equal intervals at different levels, and that positions that coincide with the start of higher-level intervals have more importance than those coinciding with lower levels.
Analogous to the perception of musical key, the perception of meter is in part related to the distribution of musical events over time~\cite{palmer90:_mental_repres_music_meter}. 

Lastly, music often transmits a sense of coherence over the course of the piece, in that it has a \emph{structural organisation} in which motifs (small musical patterns), but also larger units such as phrases, melodies or complete sections of the music are repeated throughout the piece, either literally or in an altered form.
This characteristic is reflected in the \emph{self-similarity matrix} of the music, where entry $(i, j)$ expresses the similarity between the music at positions $i$ and $j$.
This coherence by way of repeating and developing musical material over the course of the piece is arguably one of the aspects of music that make listening and re-listening a valuable experience to human listeners.

Important musical characteristics such as these are not straight-forward to capture using a naive sequence modelling approach, unless the musical material is restricted or simplified as mentioned above.
For example, it is a challenge for current models to generate music that is diverse and interesting, and at the same time induces a stable musical key over an expanded period of time.
It is even more challenging to generate music that exhibits the hierarchical organisational structure common in human-composed music.
For instance, the rather common pattern of a melodic line from the opening of a piece being repeated as the conclusion of that piece is very hard to capture, even if state-of-the-art sequence models like LSTMs are capable of learning long-term dependencies in the data.
Models that fail to capture these higher-level musical characteristics may still produce music that on a short time scale sounds very convincing, but on longer stretches of time tends to sound like it wanders aimlessly, and misses a sense of musical direction.

In this work, we do not address the problem of learning the discussed properties from musical data.
Rather, our contribution is a method to enforce such properties as constraints in a sampling process.
We start from the observation stated above, that neural network models in the various forms that have recently been proposed are adequate for learning the \emph{local} structure and coherence of the musical surface, that is, the \emph{musical texture}.
The strategy we propose here uses such a neural network (more specifically, a Convolutional Restricted Boltzmann Machine, see Section~\ref{sec:rbm}) as one of several components that jointly drive an iterative sampling process of music generation.
This model is trained on musical data, and is used to ensure that the musical texture is similar to that of the training data.

The other components involved in the sampling process are cost functions that express how well higher-level constraints like tonal, metrical and self-simi\-larity structure are satisfied in the musical material at each stage in the process.
By performing \emph{gradient descent} on these cost functions the sampling process is driven to produce musical material that better satisfies the constraints.
The desired shape of these higher-level structural constraints on the piece is not hard-coded in the cost-functions, but is instantiated from an existing piece.
As such, the existing piece serves as a \emph{structure template}.
The generation process then results in a re-instantiation of that template with novel material.
Through recombination of structural characteristics from a musical piece that is not part of the neural network's training data, the model is forced to produce novel solutions.

We refer to the above process as \emph{constrained sampling}.
Informally, it can be imagined as a musical drawing board that is initially filled with random pitches at random times, and where the neural network model as well as each of the constraints take turns to slightly tweak the current content of the drawing board to their liking.
This process continues until the musical content can no longer be tweaked to better satisfy the model and constraints jointly.

We believe this approach provides a novel and useful contribution to the problem of polyphonic music generation.
Firstly, it takes advantage of the strengths of state-of-the-art deep learning methods for data modelling.
The combination with multi-objective constraint optimisation compensates for the weaknesses of these methods for music generation, mentioned above.
Moreover, it provides high-level user control\footnote{The user has control over the generation process by the choice of the template piece, but also more directly by manipulation of the structure templates extracted from the template piece. This aspect is beyond the scope of the current article.} over the generation process, and allows for relating the generated material to existing pieces, both of which are interesting from a musical point of view.

In addition to the description of the constrained sampling approach to music generation, the goal of the present paper is to validate the approach in several ways.
First we present a qualitative discussion of generated musical samples, illustrating the effect of the constraints on the musical result.
We show that although the constraints and the neural network embody different objectives, the evolving musical material produced by the constrained sampling process tends to simultaneously approximate meeting these different objectives.
Furthermore, we adopt \emph{Information Rate} as an independent measure of musical structure \cite{wang2015pattern}, in order to assess the effect of the repetition structure constraint, and compare our approach to two variants of the state-of-the-art RNN-RBM model for polyphonic music generation \cite{boulanger2012modeling}.
This comparison shows that the constrained sampling approach substantially increases the Information Rate of the produced musical material over both unconstrained approaches (including the RNN-RBM variants), implying a higher degree of structure.

The paper is structured as follows:
Section~\ref{sec:related-work} gives an overview of related models and computational approaches to music generation.
Section~\ref{sec:method} describes the components involved in the constrained sampling approach, which is subsequently introduced in Section~\ref{sec:cs}.
Section~\ref{sec:experiment} describes the experimental validation of the constrained sampling approach in the context of Mozart piano sonatas.
We discuss the empirical findings in Section~\ref{sec:results-discussion} and give conclusions and future perspectives in Section~\ref{sec:future}.

\section{Related work}\label{sec:related-work}

Early attempts using neural networks for music generation were reported in \cite{Todd:1989bb},
where monophonic melodies were encoded in pitch and duration and an RNN was trained to predict upcoming events.
In \cite{mozer1994neural}, an RNN system called CONCERT was proposed, and first systematic tests on how well local and global musical structure (e.g.
AABA) of simple melodies could be learned, were made.
In addition, chords were used to test if this facilitates the learning of higher-level structure, but the results were not convincing.
This was one of the first papers which showed the difficulties of learning structure in music.

More recently, \citet{Eck:2002:FLM:870511} trained a Long Short-Term Memory (LSTM) network (a state-of-the-art RNN variant), jointly on a single chord sequence along with several different melodies.
This is an example of a harmonic template which guides a melodic improvisation.
Chords and melody notes were separated in the input and output connections so that the model could not mix up harmony and melody notes.
That way, the LSTM could overfit on the single chord sequence and generalise on the monophonic melodies.
In a polyphonic setting, common RNNs are not suitable for generation in a random walk fashion as the distribution at time $t$ is conditioned only on the past, but it would be necessary to consider the full joint distribution also for all possible settings in $t$.

This limitation was overcome by the RNN Restricted Boltzmann Machine (RNN-RBM) model for polyphonic music generation introduced in \cite{boulanger2012modeling}, and the similar LSTM Recurrent Temporal RBM (LSTM-RTRBM) model proposed in \cite{lyu2015modelling}.
In those architectures, the recurrent components ensure temporal consistency, while the Restricted Boltzmann Machine (RBM) component is used for sampling a plausible configuration in $t$.
Our contribution lies between the before-mentioned LSTM approach where a higher-level structure is imposed by using a template, and the RNN-RBM approach, where the ability of an RBM to model low-level structure is utilised.
Further methods to constrain generated material by pre-defining voices to guide the sampling process are introduced in \cite{hadjeres2017deepbach} (based on LSTMs), and \cite{huang2017counterpoint} (based on Convolutional Neural Networks), both of which generate Bach chorales.

Another approach that uses a probabilistic model and constraints is called ``Markov constraints'' \cite{pachet2011}, which allows for sampling from a Markov chain while satisfying pre-defined hard constraints.
This is conceptually similar to our method, but we use a different probabilistic model and soft constraints.
Our method is more flexible in defining new constraints and it is of linear runtime, while Markov constraints are more costly, but also more exact.
\citet{herremans2016morpheus} use a constrained variable neighbourhood search to generate polyphonic music obeying a tension profile and the repetition structure from a template piece.
Furthermore, \citet{barbieri2011regularized} uses soft constraints to incorporate a-priori-information in a Gibbs sampling process for a User Rating Profile model.

\citet{cope1996experiments} explicitly imposes higher-level structure in a generation process.
So-called SPEAC identifiers are used to generate music in a given tension-relaxation scheme.
Another example of generating structured material is that in \cite{eigenfeldt2013evolving}, where Markov chains and evolutionary algorithms are used to generate repetition structure for Electronic Dance Music.
\citet{collins2016developing} use Markov chains together with structure schemes and explicit methods for handling transitions between repeating segments in order to generate structured music.
Similarly, \citet{whorleytransformational} use a transformational approach to generate Bach chorales, and \citet{conklin_semiotic} generates chords using Markov chains and pre-defined repetition structures.
A Hierarchical Variational Autoencoder for music generation, able to learn hierarchical tonal structure, is proposed in \cite{roberts2017hierarchical}.

A method similar to our approach is that of \citet{gatys2016image} for image style transfer.
They also use gradient descent on the input for satisfying multiple objectives (approximating a gram-matrix defining the style, as well as the initial picture defining the structure).
In contrast to our method, there is no probabilistic model involved.
Different solutions for the same objectives are merely due to the initialisation of the input with random noise.
Application of the method to music would, however, not allow for the control needed to comply with some music-specific properties like self-similarity or the somewhat strict rules of musical tonality.

Examples of connectionist generation approaches with constraints in other domains are that in \cite{Graves:2013ua} where biasing and priming is used in LSTMs to control the generation of sequences of handwritten text, and in \cite{Taylor:2006vl} where a conditional RBM is used to generate different human walking styles.
In such problems, the number of variables is fixed and lower than in music generation, and structural properties like repetition are either not a property of the data (handwritten text) or periodic (walking), whereas polyphonic music exhibits complex structure in multiple hierarchical levels.

\section{Method}\label{sec:method}
\begin{figure}[t]
\centering\footnotesize
\includegraphics[width=\textwidth]{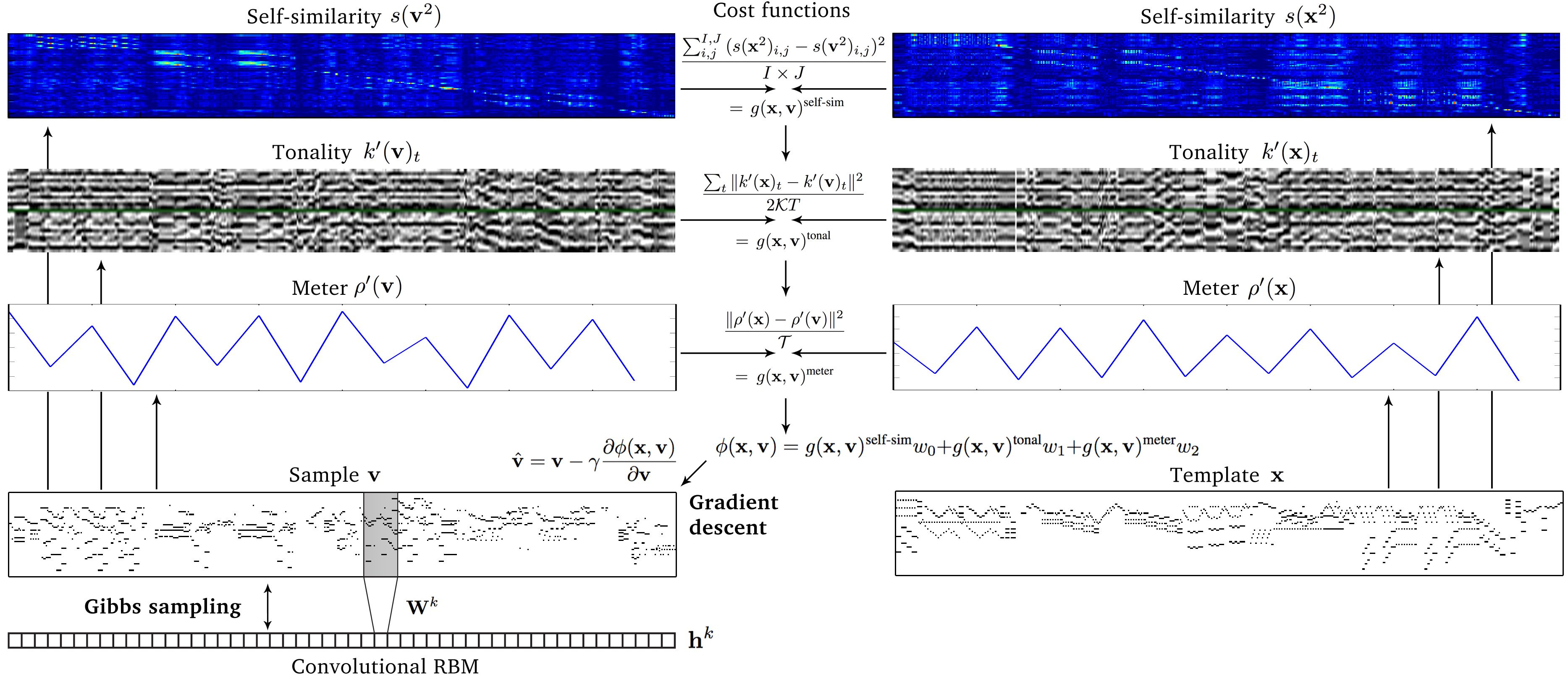}
\caption{Constrained sampling using an existing piece $\mathbf{x}$ as a structure template.
A randomly initialised sample $\mathbf{v}$ is alternately updated with Gibbs sampling (GS) and gradient descent (GD).
In GD, the error $\phi(\mathbf{x}, \mathbf{y})$ between structural features of $\mathbf{x}$ and $\mathbf{v}$ is lowered, in GS the training data distribution is approximated.
The Convolutional RBM consists of visible layer $\mathbf{v}$ and hidden layer $\mathbf{h}$.
The filter $\mathbf{W}^k$ is shared among all units in feature map $\mathbf{h}^k$.
Depicted equations are also given in Section~\ref{sec:gd}.}
\label{fig:cs}
\end{figure}

In this Section, we describe the methods used to create musical output.
We start by describing the Convolutional Restricted Boltzmann Machine used for sampling new content (Section~\ref{sec:rbm}).
The gradient descent (GD) method used to impose constraints on the sampling process is introduced in Section~\ref{sec:gd}.
The complete process, referred to as Constrained Sampling (CS) and depicted in Fig.~\ref{fig:cs}, is introduced in Section~\ref{sec:cs}.

\subsection{Convolutional Restricted Boltzmann Machine (C-RBM)} \label{sec:rbm}
\begin{figure}[t]
\centering
\includegraphics[width=.8\textwidth]{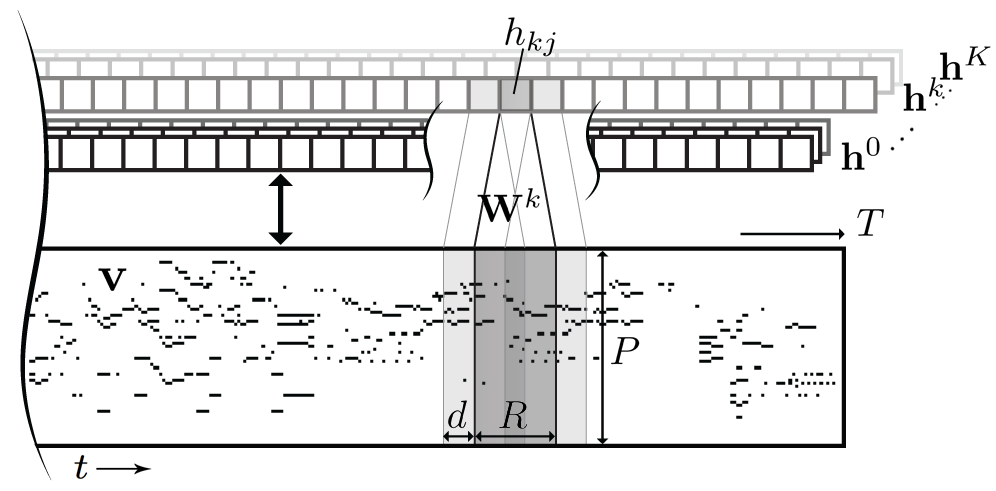}
\caption{Illustration of a C-RBM with strided convolution (using stride $d$) in the time dimension $t$ of a music piece  $\mathbf{v} \in \mathbb{R}^{T \times P}$ in two-dimensional piano roll representation using $K$ one-dimensional feature maps $\mathbf{h}^k$ where all units in a map share their weights $\mathbf{W}^k \in \mathbb{R}^{R \times P}$ and their bias $b_k$ (bias not depicted in the illustration).}
\label{fig:crbm_detail}
\end{figure}

A Convolutional Restricted Boltzmann Machine (C-RBM) \cite{lee2009convolutional}
is a two-layered stochastic version of a convolutional neural network with binary units, as known from \citet{lecun1989backpropagation}.
In our setting, the visible layer with units $\mathbf{v} \in \mathbb{R}^{T \times P}$, where $0 \leq v_{tp} \leq 1$, constitutes a piano roll representation (see Section~\ref{sec:repr}) with time $1 \leq t \leq T$ and midi pitch number $1 \leq p \leq P$.
All units in the hidden layer belonging to the $k$th feature map share their weights (i.e.
their filter) $\mathbf{W}^k \in \mathbb{R}^{R \times P}$ and their bias $b_k \in \mathbb{R}$, where $R$ denotes the \emph{filter width} (i.e.
the temporal expansion of the receptive field), and each filter covers the whole midi pitch range $[1, P]$.
We convolve only in the time dimension, which is padded with $R/2$ zeros on either side (the reason for this design decision is given in the end of this section).
We use a stride of $d$, meaning the filters are shifted over the input with step size $d$.
This results in a hidden layer $\mathbf{h} \in \mathbb{R}^{K \times (T/d)}$, where $0 \leq h_{kj} \leq 1$ and $j \in 0 \dots T/d$.
See Fig.~\ref{fig:crbm_detail} for an illustration of the C-RBM used in our experiments.

We train the C-RBM with Persistent Contrastive Divergence \cite{tielemanPcd2008} aiming to minimise the free energy function
\begin{equation}\label{equ:energy}
\mathcal{F}(\mathbf{v}) = - \sum_{t}\mathbf{a} \, \mathbf{v}^t - \sum_{k,j} \log \left( 1 + e^{\left(b_k + (\mathbf{W}^k * \mathbf{v})_{j\times d} \right)}\right)
\end{equation}

\noindent for training instances $\mathbf{v}$, where $\mathbf{a} \in \mathbb{R}^{P}$ and $\mathbf{b} \in \mathbb{R}^K$ are bias vectors, and $*$ is the convolution operator.
Note that in two-dimensional convolution, each feature map usually has a scalar as bias (e.g.
$b_k$), because all positions in a feature map are assumed to be equivalent.
However, since we convolve only in the time dimension, and since there is a non-uniform distribution over the pitch dimension, we define the bias for the input feature map $\mathbf{v}$ as a vector $\mathbf{a}$ of length $P$.

The probability of a unit being active depends on the full configuration of the opposing layer.
When updating hidden units $\mathbf{h}$ and visible units $\mathbf{v}$, each unit is randomly chosen to be active (i.e.
$1$) or inactive (i.e.
$0$) with probabilities
\begin{equation}\label{eq:ph}
P(h_{kj}=1 \mid \mathbf{v}) = \sigma \big( \big( \sum_{r,p}^{R,P}{W_{r,p}^k \times v_{j\times d+r-\frac{R}{2},p} \big) +b_k \big) }
\end{equation}
and

\begin{equation}\label{eq:pv}
P(v_{tp}=1 \mid \mathbf{h}) = \sigma\big( \big( \sum_{r,k}^{R/d,K}{\tilde{W}_{r\times d,p}^k \times h_{t+r-\frac{R}{2d}}^k\big) +a_p \big),}
\end{equation}
where $\tilde{\mathbf{W}}^k$ denotes the horizontally flipped weight matrix.
Note that it is also valid to propagate such probability values through the network (i.e.
calculate the activation probabilities of one layer based on the \emph{probabilities} of the opposing layer).

A sample can be drawn from the model by randomly initialising $\mathbf{v}$ (following the \emph{standard uniform distribution}), and running block Gibbs sampling (GS) until convergence.
To this end, hidden units and visible units are alternately updated given the other.
In doing so, it is common to sample the states of the hidden units for the top-down pass, but use the probabilities of the visible units for the bottom-up pass.
After an infinite number of such Gibbs sampling iterations, $\mathbf{v}$ is an accurate sample under the model.
In practice, convergence is reached when $\mathcal{F}(\mathbf{v})$ stabilises.

The reason for convolving only in the time dimension is that there are correlations between notes over the whole pitch range.
In a one-layered setting with 2D convolution, the filter height (i.e.
the expansion of filters in the pitch dimension) is typically limited, for example, to one octave.
In that case, correlations would only be learned between notes within one octave.
Learning correlations over a wider range would usually be the role of higher layers in a neural network stack.
However, in order to show the principle of constrained sampling it is sufficient to use only one layer with 1D convolution, which is also advantageous for limiting the overall complexity of the architecture.


\subsection{Imposing constraints with gradient descent (GD)}\label{sec:gd}
When sampling from a C-RBM, the solution is randomly initialised and converges to an accurate sample of the data distribution after many steps (see Section~\ref{sec:rbm}).
During this process, we repeatedly adjust the current solution $\mathbf{v}$ towards satisfying a desired higher-level structure regarding some musical properties.
To this end, we subject $\mathbf{v}$ (i.e.
the input, not the model parameters) to a GD optimisation process aiming to minimise a differentiable cost function $\phi(\cdot)$ using learning rate $\gamma$ as
\begin{equation}\label{equ:gd}
\hat{\mathbf{v}} = \mathbf{v} - \gamma \frac{\partial \phi(\mathbf{x},\mathbf{v})}{\partial \mathbf{v}},
\end{equation}
where $\mathbf{x} \in \mathbb{R}^{T\times P}$, $0 \leq x_{tp} \leq 1$, is a template piece from which we want to transfer some structural properties to our sample $\mathbf{v}$.
After every GD update, we set each entry
$\hat{v}_{tp} = \min(1, \max(0,\hat{v}_{t,p}))$, to ensure $\hat{\mathbf{v}} \in [0,1]^{T\times P}$.
The cost function may consist of several terms $g_d(\mathbf{x},\mathbf{v})$ (weighted with factors $w_d$), each defining a soft constraint which is to be imposed on the sample:

\begin{equation}\label{equ:cost}
\phi(\mathbf{x},\mathbf{v}) = g_0(\mathbf{x},\mathbf{v}) w_0 + \dots + g_{D-1}(\mathbf{x},\mathbf{v}) w_{D-1}.
\end{equation}

Note that $\mathbf{x}$ and $\mathbf{v}$, as representations of a musical score, could be assumed to be binary, but we define them as \emph{continuous} variables.
This is because we want to store continuous results of the GD optimisation in $\mathbf{v}$, as well as intermediate probabilities during Gibbs sampling.
Defining $\mathbf{x}$ as a continuous variable is a generalisation towards encoding note intensities or note probabilities, making it possible to express relative importance between notes.

In the following, we will introduce three constraints we tested in our experiments.
Note that the method is not limited to those constraints, and can be extended with additional terms which are differentiable with respect to $\mathbf{v}$.

\subsubsection{Self-similarity constraint}\label{sec:selfsim}
The purpose of the self-similarity constraint is to specify the repetition structure (e.g.
AABA) in the generated music piece, using a template \emph{self-similarity matrix} as a target.
Such a self-similarity representation is particularly useful, because it also provides \emph{distances} between any two parts of a piece.
Thus, the degree of similarity, including strong dissimilarity, may be encoded, too.
Such a representation abstracts from the actual musical texture and is therefore to a large extent content-invariant.
This allows for transferring the similarity structure in different hierarchical levels between pieces of different style, tonality, or rhythm.

\begin{figure}[t]
\centering
\includegraphics[width=\textwidth]{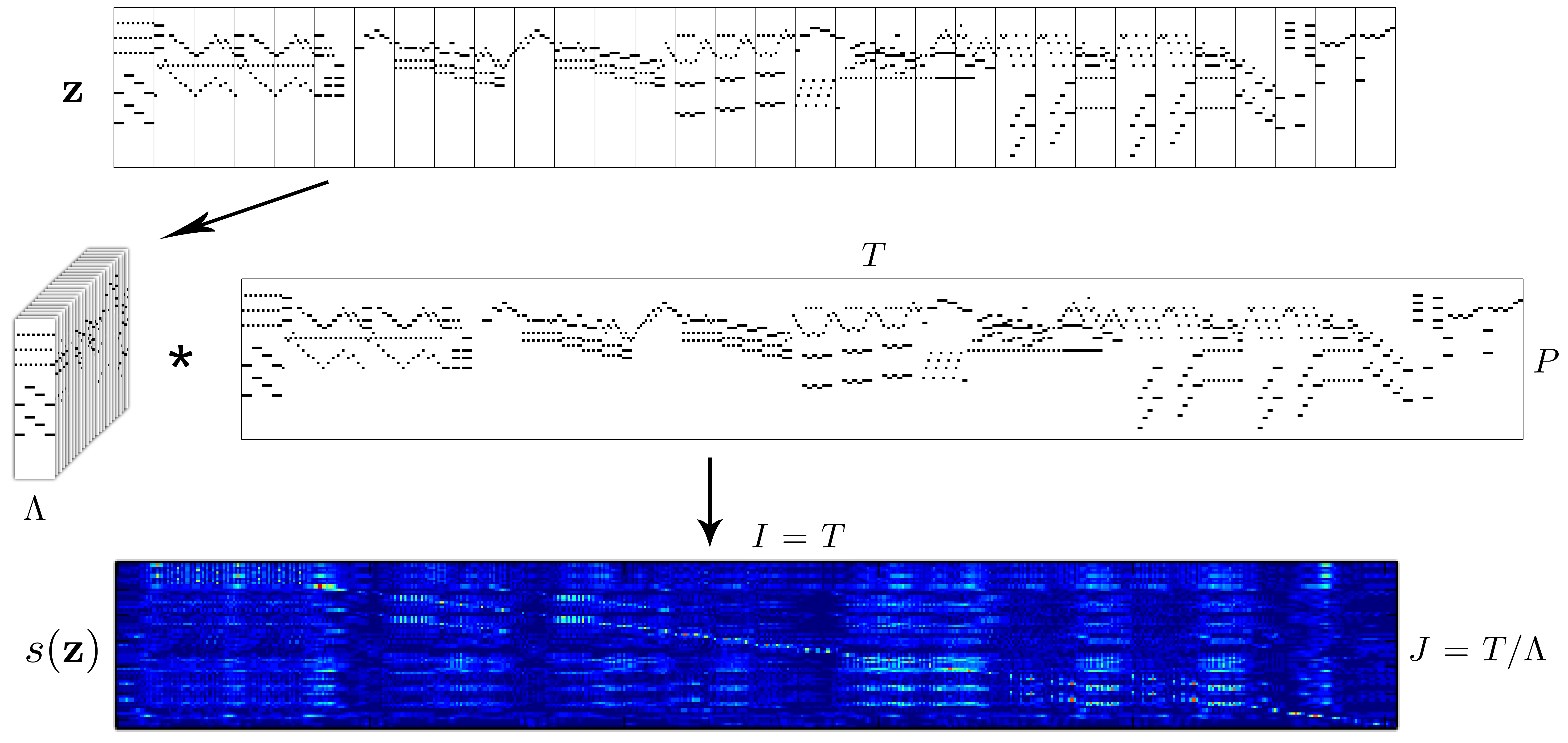}
\caption{Depiction of calculating the self-similarity matrix $s(\mathbf{z}) \in \mathbb{R}^{I \times J}$ using convolution.
A music piece in piano roll representation $\mathbf{z} \in [0,1]^{T \times P}$ is horizontally tiled, and those tiles are used as filters for a convolution with $\mathbf{z}$.
The response for a single filter constitutes a single line in the resulting self-similarity matrix.
Low to high response is depicted in a range from dark blue to light blue to red.}
\label{fig:selfsim1}
\end{figure}

A self-similarity matrix $s(\mathbf{z}) \in \mathbb{R}^{I \times J}$ for an arbitrary music piece $\mathbf{z} \in [0,1]^{T \times P}$ in piano roll representation is calculated by tiling $\mathbf{z}$ horizontally in tiles of width $\Lambda$ and by using them as 2-D filters for a convolution over the time dimension of $\mathbf{z}$ (see Fig.~\ref{fig:selfsim1}).
Therefore $I = T$ and $J = T / \Lambda$, and we calculate a single entry at position $i,j$ of the self-similarity matrix as

\begin{equation}
s(\mathbf{z})_{i,j} = \sum_{\lambda,p}^{\Lambda, P}{z_{j \times \Lambda + \lambda, p}z_{i+\lambda,p}}.
\end{equation}

To impose the self-similarity constraint, we minimise the mean squared error (MSE) between a target self-similarity matrix of the squared template piece $s(\mathbf{x}^2)$ and the self-similarity matrix of the squared intermediate solution $s(\mathbf{v}^2)$ as

\begin{equation}\label{equ:sqerrselfsim}
g(\mathbf{x},\mathbf{v})^{\text{self-sim}} = \frac{\sum_{i,j}^{I,J}{(s(\mathbf{x}^2)_{i,j} - s(\mathbf{v}^2)_{i,j})^2}}{I \times J}.
\end{equation}

The reason for squaring $\mathbf{x}$ and $\mathbf{v}$ is that it leads to a higher stability in the optimisation, because it reduces low intensity noise and it adds contrast to the resulting self-similarity matrix.
We also tried to represent transposed repetition as a constraint using two-dimensional convolution.
However, we found that this leads to a perfect reconstruction of the template piece, as such a self-similarity representation fully specifies the musical texture.


\subsubsection{Tonality constraint}\label{sec:tonality}

\begin{figure}[t]
\begin{center}
\footnotesize
\includegraphics[width=.8\linewidth]{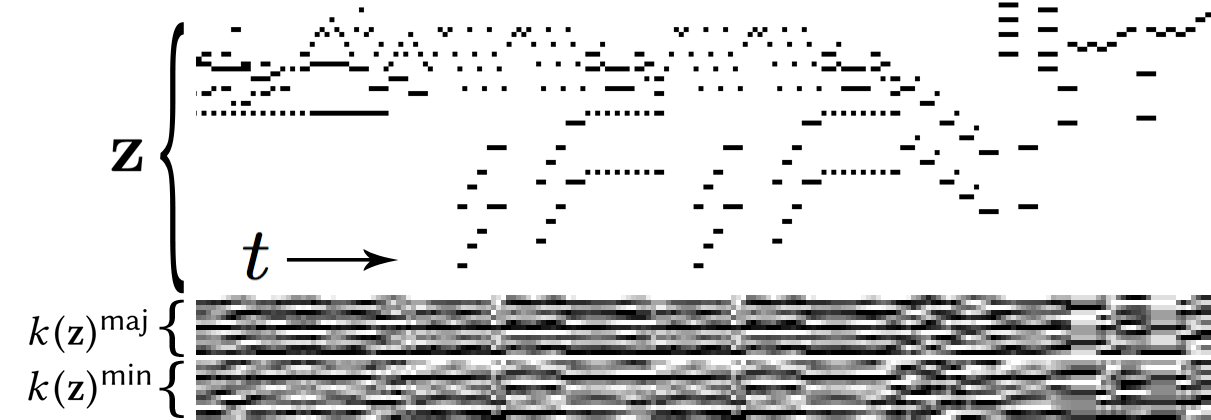}
\caption{Example of key estimation vectors over time.
$k(\mathbf{z})^\text{maj}$ represent estimations for $12$ possible major keys and $k(\mathbf{z})^\text{min}$ represent estimations for $12$ possible minor keys, where the pitch classes constituting the tonic are ordered from the top to the bottom.
Bright pixels represent high strength and dark pixels represent low strength of the respective key.
For example, the upper most line in $k(\mathbf{z})^\text{maj}$ represents the estimation strength of the C major key over time, the third line represents the strength of the D major key, etc.}
\label{fig:tonality}
\end{center}
\end{figure}

Tonality is another very important higher order property in music.
It describes perceived tonal relations between notes and chords.
This information can be used to, for example, determine the \emph{key} of a piece or a musical section.
A key is characterised by a tonal centre (the pitch class that is considered to be central, e.g. C, or A\raisebox{.45mm}{$\sharp$}), and a mode (the subset of pitch classes that form part of the key, e.g. \emph{major} or \emph{minor}).
The distribution of pitch classes in the musical texture within a (temporal) window of interest is an important factor in the perceived key of that window.
Different window lengths $M$ may lead to different key estimates, constituting a hierarchical tonal structure.
A common method to estimate the key in a given window is to compare the distribution of pitch classes in the window with so-called key profiles $u^{\text{mode}}$ (i.e.
paradigmatic relative pitch-class strengths for specific modes; the profiles are invariant to changes of tonal centre).
In \cite{temperley01}, key profiles for \emph{major mode} $u^{\text{maj}}$ and \emph{minor mode} $u^{\text{min}}$ are defined as

\begin{equation*}
u^{\text{maj}} = (5,\ 2,\ 3.5,\ 2,\ 4.5,\ 4,\ 2,\ 4.5,\ 2,\ 3.5,\ 1.5,\ 4)^\top,
\end{equation*}
\begin{equation*}
u^{\text{min}} = (5,\ 2,\ 3.5,\ 4.5,\ 2,\ 4,\ 2,\ 4.5,\ 3.5,\ 2,\ 1.5,\ 4)^\top,
\end{equation*}

\noindent where the numerical values express the strengths of the different pitch classes that make up the key.
We use these two key profiles as filters for a music piece $\mathbf{z} \in [0,1]^{T \times P}$.
By repeating them $M$ times in the \emph{time dimension}, we obtain a filter for a window of size $M$.
By repeating them $O = P/12$ times in the \emph{pitch dimension}, we extend the filters over all octaves represented in $\mathbf{z}$.
When shifted in the pitch dimension with shifts $\kappa \in 0 \dots \mathcal{K}-1$ we obtain a filter for each of the $\mathcal{K} = 12$ possible keys.
If we choose the profile for a specific mode $u^{\text{mode}}$, an estimation window size $M$, and the number of octaves $O$ represented by $\mathbf{z}$, we obtain a key estimation vector $k(\mathbf{z})_t^\text{mode} \in \mathbb{R}^\mathcal{K}$ at time $t$ for all shifts $\kappa$ as

\begin{equation}
k(\mathbf{z})_t^\text{mode} = \sum_{m,o,i}^{M,O,I}{u^{\text{mode}}((i+\kappa)\bmod I) \cdot z_{t+m,i+o*12}},
\end{equation}
for $I = 12$ entries in key profile $u^{\eta}$, where $\cdot$ denotes the common multiplication of scalars.
Subsequently, we concatenate the key estimation vectors of both modes, $k(\mathbf{z})_t^\text{maj}$ and $k(\mathbf{z})_t^\text{min}$, to obtain a combined estimation vector $k(\mathbf{z})_{t} \in \mathbb{R}^{2\mathcal{K}}$ in $t$, which is finally normalised as
\begin{equation}
k'(\mathbf{z})_{t} = \frac{k(\mathbf{z})_{t} - \min(k(\mathbf{z})_{t}) \mathbf{I}}{\max(k(\mathbf{z})_{t}) - \min(k(\mathbf{z})_{t})},
\end{equation}
where $\mathbf{I}$ is a vector of ones of length $2\mathcal{K}$.\footnote{Even though the derivatives of $\min(\cdot)$ and $\max(\cdot)$ are not guaranteed to be always defined, in practice these cases are hardly ever a problem in gradient descent, and are typically dealt with in software frameworks for symbolic differentiation such as Theano~\cite{2016arXiv160502688short}.}
Fig.~\ref{fig:tonality} depicts the resulting concatenated key estimation vectors.
Using these vectors, we may impose a specific tonal progression on our solution by minimising the MSE between the target estimate $k'(\mathbf{x})_t$ and the estimate of our current solution $k'(\mathbf{v})_t$ such that:

\begin{equation}
g(\mathbf{x},\mathbf{v})^{\text{tonal}} = \frac{\sum_t\|k'(\mathbf{x})_{t} - k'(\mathbf{v})_{t}\|^2}{2\mathcal{K}T}.
\end{equation}

\subsubsection{Meter constraint}
The meter (e.g.
$3/4$, $4/4$, $7/8$) defines the \emph{duration} and the perceived \emph{accent patterns} in regularly occurring bars of a music piece.
For example, in a $4/4$ meter, relatively strong accents on the first and the third beat of a bar are common.
We impose a common meter extracted from a template piece on our sample, to obtain a degree of global rhythmic coherence.

Perceived accent patterns depend on the relative occurrence of note onsets in a bar, on the intensity of played notes, or on the length of notes starting at the respective positions of a bar.
However, note intensities are not encoded in our data, and it is not obvious how to incorporate note durations in our differentiable cost function.
Therefore, we use note onsets only.
To this end, we constrain the relative occurrence of note onsets within a bar to follow that of a template piece.
Abiding by such a distribution helps the generated material to keep implying a regular meter.

The onset function $\omega(\cdot)$ results from a discrete differentiation over the time dimension of an arbitrary music piece in piano roll representation $\mathbf{z} \in [0,1]^{T \times P}$.
We rectify that result (as we are not interested in note offsets), and sum over the pitch dimension:
\begin{equation}
\omega(\mathbf{z},t) = \sum_p^P\max(0, {z_{t,p} - z_{t-1,p}}).
\end{equation}

In order to calculate the relative occurrences of onsets within a bar, the length $\mathcal{T}$ of a bar has to be pre-defined.
We count the number of onsets occurring on the respective positions of all bars in the music piece.
That is, we sum up all values of distance $\mathcal{T}$ in the onset function $\omega(\cdot)$ as
\begin{equation}
\rho(\mathbf{z})_{\tau} = \sum_{\mu}^{T/\mathcal{T}}{\omega(\mathbf{z}, \tau + \mu*\mathcal{T})},
\end{equation}
where $\tau \in 0 \dots \mathcal{T}-1$ is the position in a bar.
In our experiments, we use a resolution of $16th$ notes in the representation and the template is in $4/4$ meter, therefore $\mathcal{T} = 16$.

To keep the function independent of the absolute number of onsets involved, $\rho(\mathbf{z})$ is standardised by subtracting its mean $\overline{\rho(\mathbf{z})}$ and dividing through its standard deviation $\sigma(\rho(\mathbf{z}))$, resulting in zero mean and unit variance:
\begin{equation}\label{equ:onsetdist}
\rho'(\mathbf{z}) = \frac{\rho(\mathbf{z}) - \overline{\rho(\mathbf{z})}}{\sigma(\rho(\mathbf{z}))}.
\end{equation}

\begin{figure}[t]
\begin{center}
\footnotesize
\includegraphics[width=.8\linewidth]{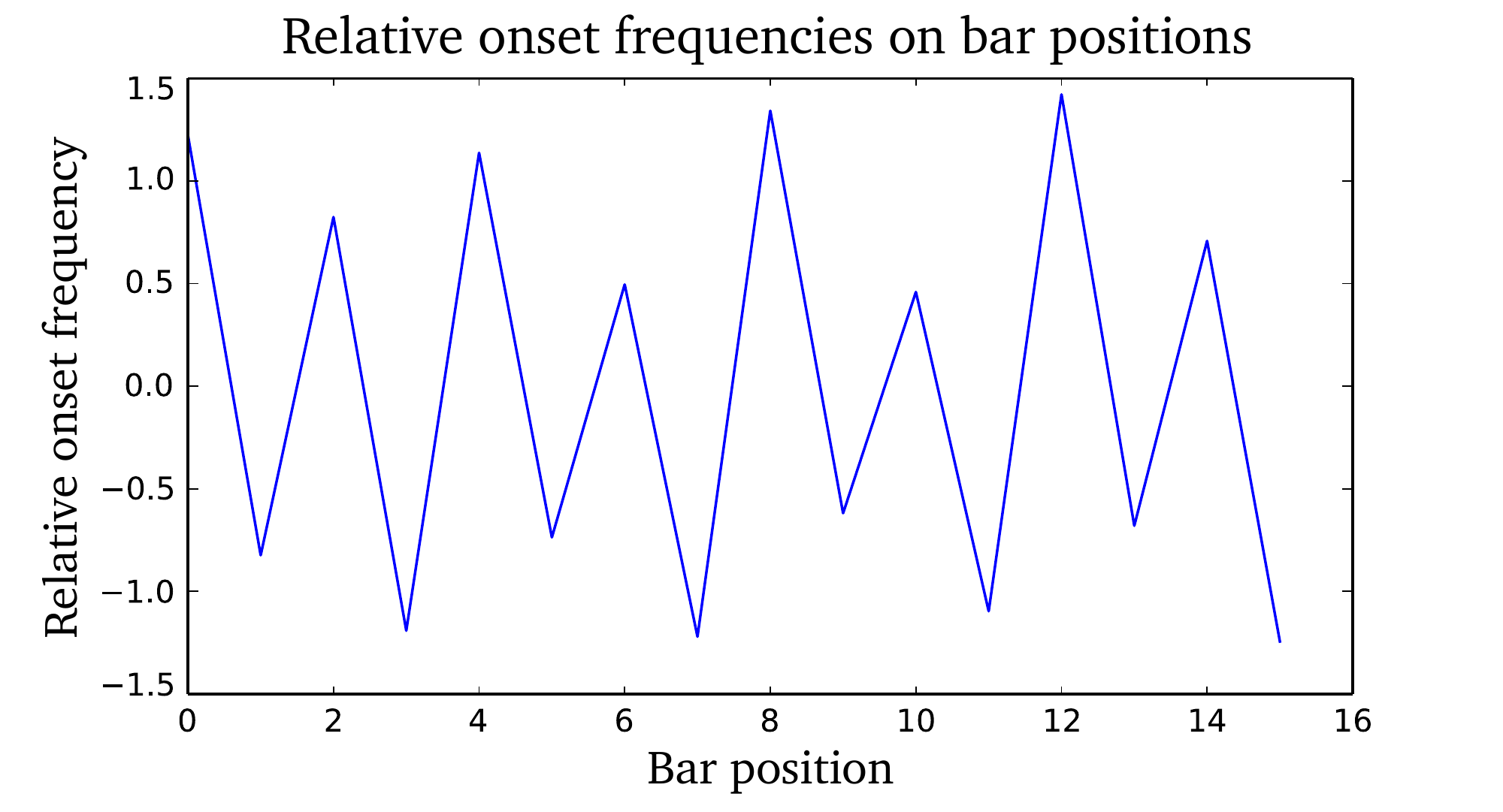}
\caption{Relative (standardised) onset frequencies $\rho'(\mathbf{z})$ on bar positions of a music piece as obtained from Equation~\ref{equ:onsetdist}.}
\label{fig:onsets}
\end{center}
\end{figure}

A standardised onset distribution is plotted in Fig.~\ref{fig:onsets}.
Finally, we minimise the MSE between a standardised onset distribution $\rho'(\mathbf{x})$ and that of our intermediate solution $\rho'(\mathbf{v})$ as

\begin{equation}
g(\mathbf{x},\mathbf{v})^{\text{meter}} = \frac{\|\rho'(\mathbf{x}) - \rho'(\mathbf{v})\|^2}{\mathcal{T}}.
\end{equation}

\section{Constrained Sampling}\label{sec:cs}



In this Section, we describe how the C-RBM is used as a generative model to produce musical textures that resemble those of human-composed music, and combined with the soft constraints described above, to enforce additional tonal, meter and self-similarity structure on these textures.





\begin{sidewaysfigure}
\centering
\includegraphics[width=1\textwidth]{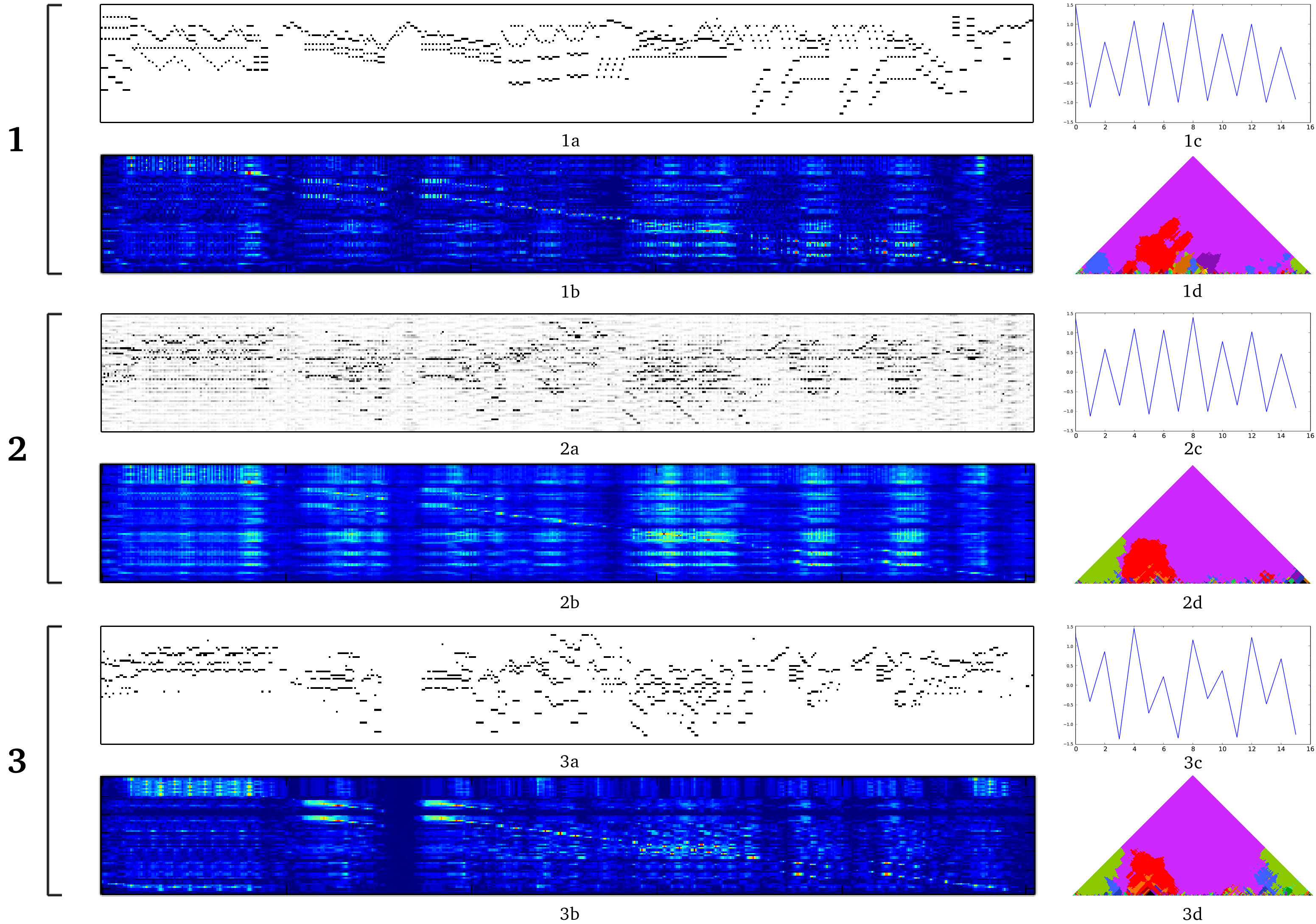}
\caption{Illustration of constrained sampling.
(1) Template piece, (2) Intermediate sample after the GD phase, (3) Sample after the GS phase.
Fig.s in each group: (a) Piano roll representation, (b) Self-similarity matrix, (c) Onset distribution in $4/4$ meter, (d) Keyscape (cf. Section \ref{sec:keyscape} for an explanation).
After the GD phase (2), the target higher-level properties imposed as constraints are relatively well approximated.
Due to limited training data and the stochastic nature of Gibbs sampling, after the GS phase the higher-level properties are more dissimilar again.}
\label{fig:iteration}
\end{sidewaysfigure}

The method proposed here has several practical merits.
First, a C-RBM can take any input as a starting point for (further) sampling.
This allows for local ``mutations'' of intermediate solution candidates in a heuristic process like Simulated Annealing (see Section~\ref{sec:simulatedannealing}) and facilitates the controlled exploration of the search space.
Second, in a C-RBM continuous values in the input are interpreted as probabilities.
This facilitates external guidance through gradual adaptation of note probabilities in a directed gradient descent (GD) optimisation process.
For illustration, Fig.~\ref{fig:iteration}[2a] shows an example of a piano roll after a GD phase.
The grey tones (non-zero probabilities) in the background of the piano roll will influence the subsequent sampling step from the C-RBM.  
Third, the solution is sampled as a single instance (i.e.
all notes in a music piece are updated simultaneously) and temporal dependencies are modelled in a bi-directional manner.
That way, global constraints can be imposed by iterative adaptation of local structures.

\subsection{Example scheme and details}

In Fig.~\ref{fig:cs}, an overview of Constrained Sampling (CS) is shown.
During the constrained sampling process we alternate between a GS phase with the one-layered C-RBM (Section~\ref{sec:rbm}), and a GD optimisation phase on the cost functions (Section~\ref{sec:gd}).
In each phase, typically multiple iterative updates take place, and the sampling results are sensitive to the balance struck between the GS and GD phases, in terms of the number of updates performed in each phase.

The numbers proposed in the following CS sampling scheme have been found to work well in our experiments.
The scheme may have to be adapted to work well with other training settings (e.g. different C-RBM architectures, or different constraints), and is mainly for illustrative purposes.
In Algorithm~\ref{alg:cs}, the whole process including Simulated Annealing (see Section~\ref{sec:simulatedannealing}) is shown.

\begin{algorithm}[t]
\KwData{

$\mathbf{x} \in [0,1]^{T \times P}$ -- Template Piece

$\mathbf{v} \in [0,1]^{T \times P}$ -- Random (standard uniform dist.)
$\mathbf{\hat{v}} = \mathbf{v}, N = 250, M = 15$
}

\For{$i \in 1 \dots N$}{
$\mathbf{v}' \leftarrow \mathbf{v}$

$\mathbf{v} \leftarrow$ 20 GD steps using Eq.~\ref{equ:gd} with $\gamma = 10$

\For{$j \in 1 \dots M$}{
$\mathbf{v} \leftarrow$ 100 GS steps using $\mathbf{v}$

$\mathbf{v} \leftarrow$ 1 GD step using Eq.~\ref{equ:gd} with $\gamma = 5$
}
\tcc{Simulated Annealing}
$T_i = 1 - i/N$

$r_e, r_c \leftarrow$ random values $\in [0,1]$

\If {$r_e < \exp\left(-\frac{\mathcal{F}'(\mathbf{v}) - \mathcal{F}'(\mathbf{v}')}{T_i}\right)$ \textbf{or} $r_c < \exp\left(-\frac{\phi'(\mathbf{x},\mathbf{v})- \phi'(\mathbf{x},\mathbf{v}')}{T_i}\right)$}{
$\mathbf{v} \leftarrow \mathbf{v}'$
}
\tcc{Store best solution so far}
\If {$\frac{\mathcal{F}'(\mathbf{v}) + \phi'(\mathbf{x},\mathbf{v})}{2} < \frac{\mathcal{F}'(\mathbf{\hat{v}}) + \phi'(\mathbf{x},\mathbf{\hat{v}})}{2}$}
{$\mathbf{\hat{v}} \leftarrow \mathbf{v}$}
}
\Return
$\mathbf{\hat{v}}$
\caption{Constrained Sampling.
Number of iterations represent an example scheme, as used in the experiments.} \label{alg:cs}
\end{algorithm}

Starting from a random uniform noise in $\mathbf{v}$, we alternate $20$ GD steps using learning rate $\gamma = 10$ (i.e.
GD phase, see Fig.~\ref{fig:iteration}[2a] for a result of this phase), and $1500$ GS steps (i.e.
GS phase, see Fig.~\ref{fig:iteration}[3a] for a result of this phase).
We consider this \emph{one} constrained sampling iteration.
We found that results improve when, during the GS phase, after every $100$ GS steps we execute $1$ GD step with learning rate $\gamma = 5$.
After 250 CS iterations, the sample with the minimal average value of the standardised GD cost and the standardised free energy over the whole CS process is chosen (see Section~\ref{sec:simulatedannealing} on standardising the cost and free energy functions).

\begin{figure}[t]
\centering\footnotesize
\includegraphics[width=.6\linewidth]{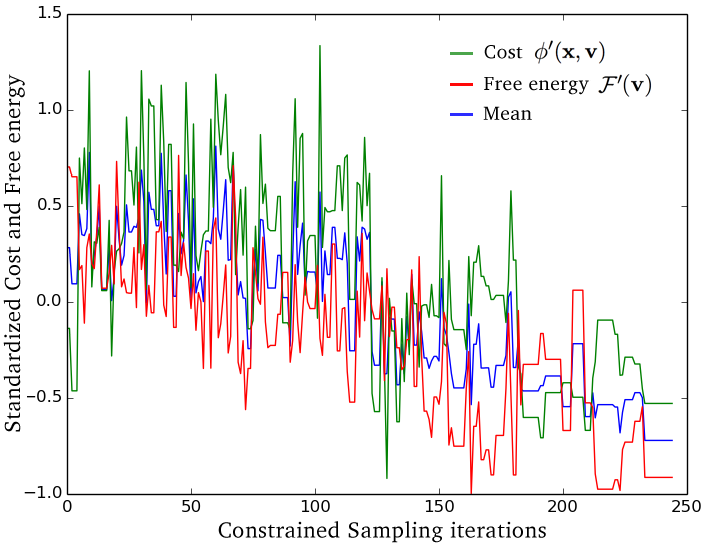}
\caption{Standardised cost, free energy and their mean in a constrained sampling process over 250 iterations.
Periods of constant cost (horizontal line segments) in later iterations are a result of Simulated Annealing, where some unfavourable solution candidates are rejected.}
\label{fig:cost_curve}
\end{figure}

\begin{figure}[t]
\centering\footnotesize

\includegraphics[width=.7\linewidth]{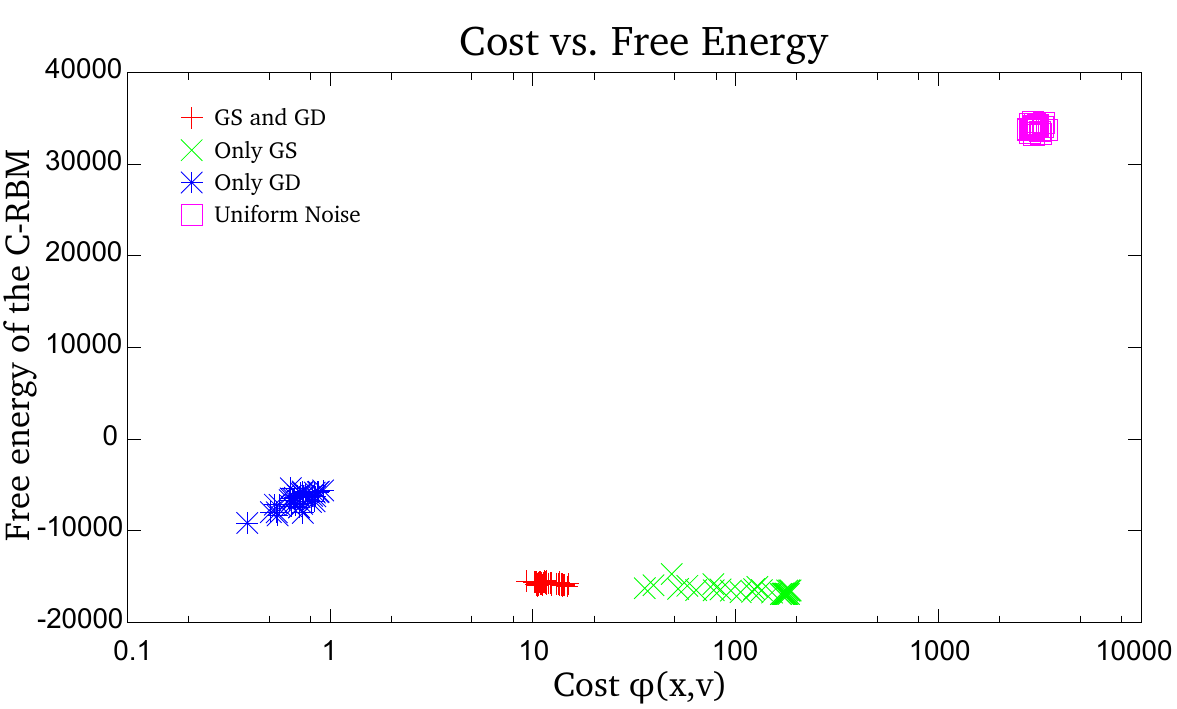}

\caption{Influence of Gibbs sampling (GS) and gradient descent (GD) on free energy $\mathcal{F}(\mathbf{v})$ (see Equation~\ref{equ:energy}) and cost $\phi(\mathbf{x},\mathbf{v})$ (see Equation~\ref{equ:cost}).
Using only GS results in low free energy but relatively high cost.
Using only GD, the cost is very low but the free energy is high.
When using GS and GD, both methods compete, resulting in a trade-off between low cost and low free energy although we choose enough GS steps in the GS phase to always return to a ``meaningful'', low free energy state.
For reference we test against random uniform noise, resulting in very high free energy and cost.
For each cluster, 50 data points were generated with the trained C-RBM model (for GS) and the cost function (for GD) used in our experiment (see Section~\ref{sec:experiment})}.
\label{fig:costvsfe}
\end{figure}

During CS, in the C-RBM the free energy is to be reduced (i.e.
a high probability solution is to be found), while in GD optimisation the objective function is to be minimised (see Fig.~\ref{fig:cost_curve} for a plot of the curves).
As the two models used compete in approximating their objectives (see Fig.~\ref{fig:costvsfe}), their mutual influence has to be balanced.
In addition to using Simulated Annealing to prevent strong deteriorations of the solution with respect to the objectives (see Section~\ref{sec:simulatedannealing}), some parameters need to be carefully adjusted.

The main parameters for balancing the models are the number of GD and GS steps used in a CS iteration, as well as the learning rate and the relative weighting of the cost terms in the GD optimisation (see Tab.~\ref{tab:weights} for weightings used in our experiments).
In general, the optimal number of steps in each model is inversely proportional to the size of the training corpus.
The more training data, the more possible solutions can be sampled by the probabilistic model making it easier to satisfy constraints imposed by the GD optimisation.
Conversely, with a model trained on little data, more GS steps are necessary in order to find another low free energy solution after being distracted by the GD phase.

Although we do not provide a formal convergence proof, all experiments show a joint decrease of the various quantities to be minimised (the C-RBM free energy, and the cost-functions of each of the constraints). Convergence is reached when both the gradient descent cost and the free energy of the C-RBM reach a minimum. When reaching equilibrium in an RBM, the visible unit configuration (the sample) keeps changing during further sampling while the free energy remains at the minimum. Therefore, with our method convergence is reached with respect to the overall (average) cost, but not with respect to a final solution in v.

\subsection{Simulated Annealing (SA)}\label{sec:simulatedannealing}
Due to the interdependency between the sampling process of the probabilistic model and the GD optimiser, it can easily happen that good intermediate solutions deteriorate again by further sampling.
Simulated Annealing (SA) helps to find good minima by preventing sampling steps which would lower the solution quality too much (see Algorithm~\ref{alg:cs} for the integration of SA in CS).
After each constrained sampling (CS) iteration, we evaluate the SA equation to obtain the probability
\begin{equation}
p_k(\mathbf{v},\mathbf{v}',i) = \exp\left(-\frac{f(\mathbf{v}) - f(\mathbf{v}')}{T_i}\right)
\end{equation}
of keeping solution candidate $\mathbf{v}$, where $\mathbf{v}'$ is the previous solution.
We evaluate this equation twice after each CS iteration.
The first time, $f(\boldsymbol{\cdot})$ is the \emph{standardised} RBM free energy function $\mathcal{F}'(\boldsymbol{\cdot})$ (see Equation~\ref{equ:energy}) and the second time, $f(\boldsymbol{\cdot})$ is the \emph{standardised} GD cost function $\phi'(\boldsymbol{\cdot})$ (see Equation~\ref{equ:cost}).
For each of the two resulting probabilities we generate a random number between $0$ and $1$, and evaluate if it is smaller than the respective probability.
If this is the case for both random numbers, we go on with solution candidate $\mathbf{v}$, otherwise we return to the former solution $\mathbf{v}'$.
The most important factor for the sensitivity of SA is the variance of $f(\boldsymbol{\cdot})$ over all solutions, where a higher variance leads to smaller probabilities for acceptance of a solution.
Therefore, we standardise $\mathcal{F}(\boldsymbol{\cdot})$ and $\phi(\boldsymbol{\cdot})$, resulting in $\mathcal{F}'(\boldsymbol{\cdot})$ and $\phi'(\boldsymbol{\cdot})$, to obtain comparable probabilities in SA.
This is done by scaling both functions to approximately zero mean and unit variance, based on the observed values during the experiments.
As the annealing scheme we use $T_i = 1 - i/N$.
In Fig.~\ref{fig:cost_curve} the standardised curves over a CS process are depicted.
In later iterations, Simulated Annealing causes periods of constant cost, as some solution candidates are rejected.

\begin{table}
\begin{center}
\begin{tabular}{@{}ll@{}}
\toprule
Constraint & $w_d$ \\ \midrule
Self-similarity & 1.5 \\
Tonality & 5.0 \\
Meter & 0.5 \\
\bottomrule
\end{tabular}
\vspace{4mm}
\caption{Relative weightings $w_d$ of the terms used in the GD objective function $\phi(\mathbf{x},\mathbf{v})$ (see Section~\ref{sec:gd}).}
\label{tab:weights}
\end{center}
\end{table}




\section{Experiment}\label{sec:experiment}
This section describes an experimental validation of the method described in Sections~\ref{sec:method} and \ref{sec:cs}.
In Section~\ref{sec:corpus} and Section~\ref{sec:repr}, we introduce the training data and the data representation scheme, respectively.
Section~\ref{sec:training} describes the training of the C-RBM.
Section~\ref{sec:qu_evaluation} introduces a quantitative measure for the structural organisation of music, adopted from \cite{wang2015pattern}.
We use this measure to evaluate the success of the constrained sampling approach, both with respect to original musical data, and with respect to other state-of-the-art polyphonic music generation models.
Lastly, Section~\ref{sec:procedure} briefly describes the procedure followed to produce musical material for qualitative evaluation.



\subsection{Training Data}\label{sec:corpus}
We use MIDI files encoding the scores of the second movement of three Mozart piano sonatas, as encoded in the Mozart/Batik data set~\cite{widmer2003discovering}: Sonata No. 1 in C major, Sonata No. 2 in F major and Sonata No. 3 in B flat major.
When applying a (major) tonality constraint, we want to make sure that there is enough training data for the probabilistic model in any possible (major) key.
Otherwise, in the GS phase, an intermediate solution might be always changed back from a key imposed by the GD optimisation to the closest key available in the training data.
Therefore, we transpose each piece into all possible keys, which also helps to reduce sparsity in the training data.
This results in a training corpus size of 15144 time steps (of sixteenth note resolution).

\subsubsection{A note on training data set size}
A widely shared insight in machine learning is that to train neural network models effectively, more data is better.
In general, it is easy to see why this is the case, since larger amounts of data provide a richer coverage of the relations to be learned in the data.
However, depending on the intended purpose of the model, there may be exceptions to this rule.
In the present study, where the main purpose of the model is to generate plausible musical textures in the style of Mozart piano sonatas, we have found that the set of all available training data (34 pieces) is likely too small for the C-RBM model to approximate the data distribution well enough to produce samples of high musical quality.
A pragmatic trade-off we have chosen in this case is to reduce the size of the training data to a few pieces, and let the model slightly overfit those pieces.
This will improve the musical quality of the samples, at the cost of increased local resemblances of generated samples to fragments of the training data.

\subsection{Data Representation}\label{sec:repr}
We transform MIDI data in a binary piano roll representation of $T=512$ time steps over a range of $P=64$ pitches (MIDI pitch number $28$-$92$), using a temporal resolution of sixteenth notes (see Fig.~\ref{fig:iteration}[1a]).
Notes are represented by active units (black pixels), and note durations are encoded by activating units up to the note offset.
If two notes directly follow each other at the same pitch, they cannot be distinguished any more.
Thus, the first note is shortened by a sixteenth note if possible (i.e.
if it is longer than a sixteenth note), otherwise the merger has to be accepted.
Note that a temporal subdivision of sixteenth notes cannot represent all rhythmic patterns in the data without distortion.
For example, the durations \{1/12, 1/12, 1/12\} of 1/8 note triplets (as contained in the Sonata No. 1) change to \{1/16, 1/8, 1/16\} using this representation.
We accept this bias, as it does not hinder our efforts to test the influence of constraints on a generated texture.

\subsection{Training}\label{sec:training}
We train a single C-RBM using Persistent Contrastive Divergence (PCD)~\cite{tielemanPcd2008} with $10$ fantasy particles, using learning rate $15 \times 10^{-4}$.
Compared to standard Contrastive Divergence~\cite{hinton06}, the PCD variant is known to draw better samples.
One training instance has a length $T = 512$, and we use a batch size of $1$.
The \emph{filter width} R (see Section~\ref{sec:rbm}) is set to 17, and we convolve only in the time dimension with stride 4, using 2048 hidden units.

We apply the well-known L1 and L2 weight regularisation with strengths $8 \times 10^{-4}$ and $1 \times 10^{-2}$, respectively, to prevent overfitting and exploding weights.
In addition, we use the max-norm regularization \cite{srebro2005rank}, which is an additional protection against exploding weights when using high learning rates.
We use sparsity regularization as introduced in \cite{Lee:2007uz}, to increase sparsity and selectivity in the hidden unit activations, leading to a better generalization of the data.
When training with PCD, it can happen that single neurons are always active, independent of the presented input.
Therefore we reset (i.e.
randomize) the weights of any neuron which exceeds the threshold of $0.85$ average activation over the data.

\subsection{Quantitative Evaluation}\label{sec:qu_evaluation}
Based on the observation that probabilistic models can generate meaningful low-level structure but struggle in obeying some higher-level structure, the focus of this study is to increase the structural organization of the generated material. In the important case of self-similarity structure, a critical property in music is the balance of repetition and variation. This ratio is expressed by an information theoretic measure called Information Rate (IR). It is the mutual information between the present and the past observations and is maximal when repetition and variation is in balance. Thus, the IR is minimal for random sequences, as well as for very repetitive sequences. It has been shown that it provides a meaningful estimator on musical structure, for instance in parameter selection for musical pattern discovery \cite{wang2015pattern}.

For a given sequence $v_0^N = \{v_0,v_1,v_2,\dots,v_N\}$, the average IR is defined by
\begin{equation}
IR(v_0^N) = \frac{1}{N}\sum_n^N{H(v_n) - H(v_n \mid v_0^{n-1})},
\end{equation}
where $H(v)$ is the entropy of $v$, which is estimated based on the statistics of the sequence \emph{up to event} $v_n$.
We approximate $H(v_n \mid v_0^{n-1})$ using a first-order Markov Chain, and $H(v_n)$ by counting identical time slices.
It may seem counter-intuitive at first sight to utilise a first-order model for measuring the higher-level structure of a piece.
Arguably, a low-order estimation yields too optimistic IR values, as the conditional entropy tends to be underestimated.
However, the initial idea of contrasting the prior entropy of events with their conditional entropy is still applicable using a first-order entropy estimation.
That is, a high IR is achieved when specific events occur rarely, but are very likely given their direct predecessors -- a situation which occurs particularly in sequences with higher-level repetition structure.
Note that the IR does not provide a measure for the overall musical quality of the evaluated sequences, but only for the aspect of self-similarity structure from an information theoretic point of view.

\subsubsection{Model comparison}
In addition to using the C-RBM without constraints, we use the RNN-RBM \cite{boulanger2012modeling}, a state-of-the-art polyphonic music generation model, as a baseline for the quantitative evaluation. Furthermore, we replace the RNN portion of the RNN-RBM with Gated Recurrent Units \citep[GRUs,][]{cho2014learning} resulting in a GRU-RBM. Both recurrent models are trained on the same data as the C-RBM, as described in Section \ref{sec:corpus}\footnote{Samples from the RNN-RBM and the GRU-RBM can be listened to on Soundcloud under \url{http://www.soundcloud.com/pmgrbm}}.

\begin{figure}[t]
\centering\footnotesize
\includegraphics[width=\linewidth]{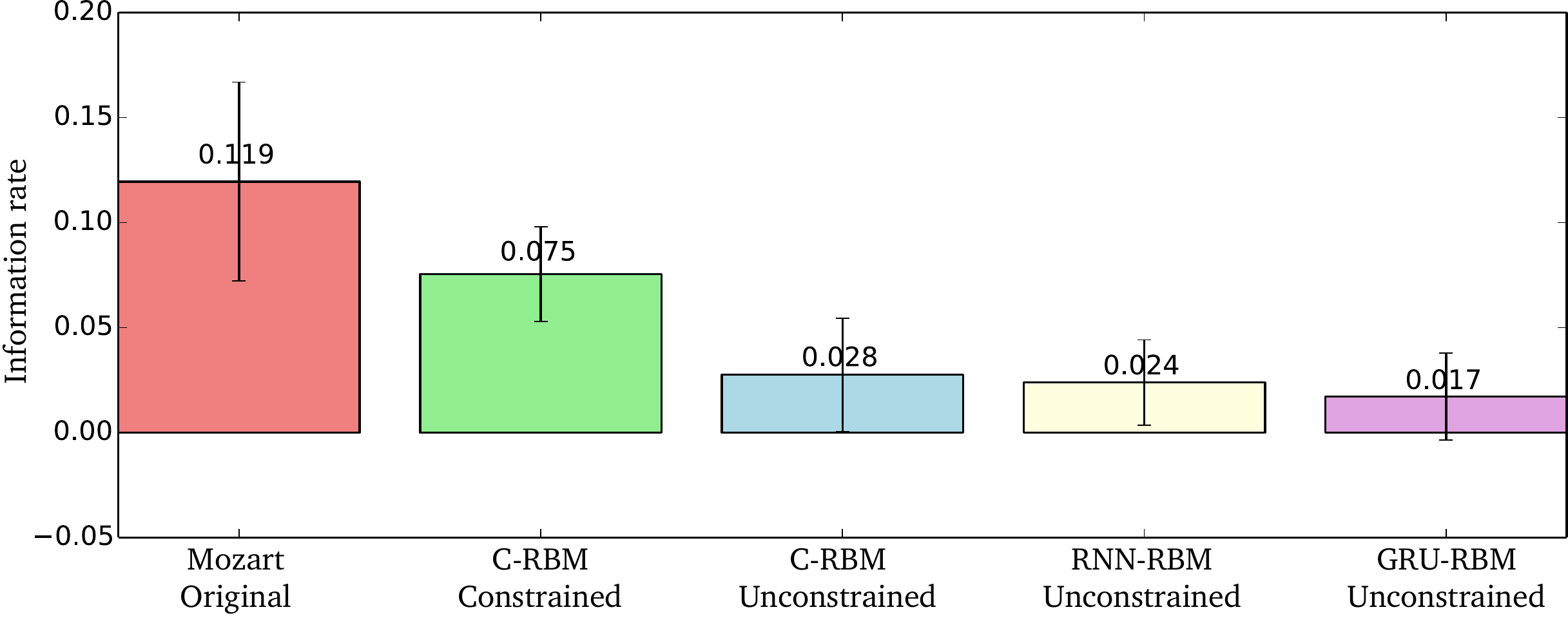}
\vspace{-0.2cm}
\caption{Box plot showing Average Information Rates for $34$ original Mozart piano sonatas, $102$ C-RBM samples with structure constraints, $102$ C-RBM samples without constraints, $102$ samples from an RNN-RBM and $102$ samples from a GRU-RBM.
Whiskers show standard deviations.}
\label{fig:info_rate}
\end{figure}

We compare the average Information Rates between original Mozart piano sonatas (all $34$ pieces of the Mozart/Batik data set, Widmer, 2003), C-RBM constrained samples using the original pieces as structure templates (3 samples per original piece resulting in $102$ samples with different lengths), $102$ C-RBM unconstrained samples, $102$ RNN-RBM unconstrained samples and $102$ GRU-RBM unconstrained samples. The $102$ unconstrained samples per model are created by generating three samples for each original piece of the length of the original piece. For the results of this comparison see Fig.~\ref{fig:info_rate}, for a discussion see Section \ref{sec:results-discussion}.

\subsubsection{Further measures for evaluating musical structure}
Information theory could provide additional quantitative measures for the evaluation of structure in music.
An important basis for that is Information Content (IC), a measure of the predictability of an event in a specific context.
Prior research has shown that IC can act as a kernel for determining segment boundaries \cite{Pearce:2010ig,lattner2015probabilistic}.
Evaluating the plausibility of IC over time in generated sequences could therefore constitute an adequate measure for the evaluation of musical structure.

Another useful theory, the principle of uniform information density (UID), originates in linguistics.
It is based on the proposal of  \citet{shannon2001mathematical} that for optimal data flow through a noisy channel, the transferred information density (i.e. the IC per time step) should be as uniform as possible.
It was shown that speakers intuitively follow these rules to keep the processing effort of the receiver at a moderate level \cite{jaeger2007speakers, aylett2006language}.

Recent research provides evidence that UID could also account for structural decisions in music composition, where it implies that the average IC in any window of fixed duration over a musical piece should be constant.
For example, \citet{temperley2014information} states ``There is a tendency that when an intervallic pattern is repeated with alterations, the alterations tend to lower the probability of the pattern rather than raising it''.
When a musical passage is repeated, its Information Content (i.e. the listeners surprise) declines.
The finding mentioned above provides some evidence that in such cases, surprising alterations should be inserted to keep the UID constant.

Since the IR is sufficient for evaluating our results, we do not use IC and UID.
Nevertheless, they seem promising as further evaluation measures to quantify structure in generated music.

\subsection{Qualitative Evaluation}\label{sec:procedure}
\begin{figure}
\centering\footnotesize
\includegraphics[width=1.\linewidth]{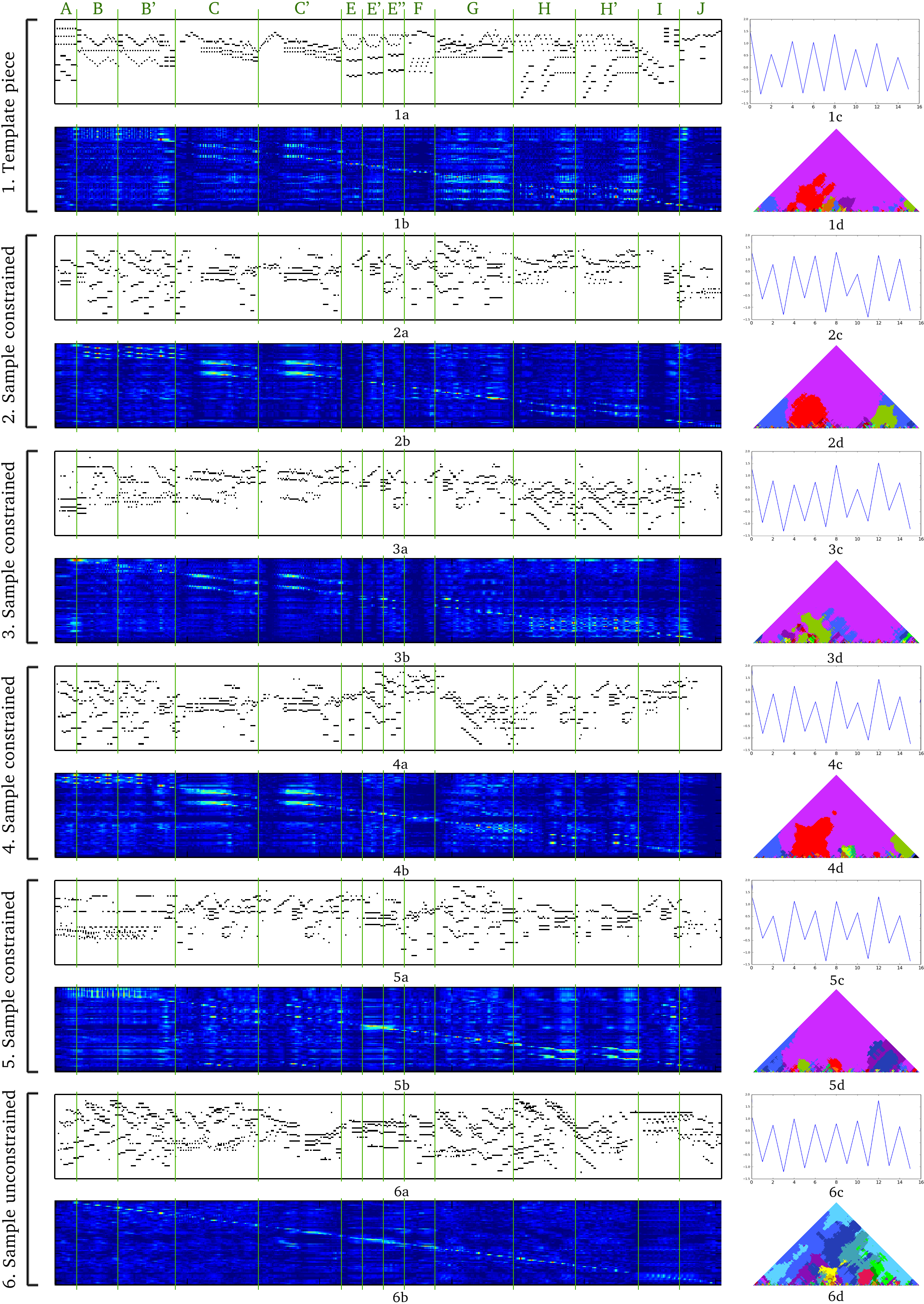}
\vspace{-0.2cm}
\caption{Template piece (1), Constrained samples (2 to 5) and an unconstrained sample as baseline (6).
Fig.s in each group: (a) Piano roll representation, (b) Self-similarity matrix, (c) Onset distribution in $4/4$ meter, (d) Keyscape.
By constrained sampling, the template piece's self-similarity and tonal structure, as well as the onset distributions, are transferred to the generated solutions 2 to 5.
The unconstrained sample (6) at the bottom was sampled without constraints, and thus does not reflect the structure of the template piece.}
\label{fig:selfsim}
\end{figure}

The C-RBM is trained as described in Section~\ref{sec:training} on the Mozart Sonatas (see Section~\ref{sec:corpus}).
After that, we pick a template piece (the first movement of the piano sonata No.
6 in D major) and generate constrained samples, as introduced in Section~\ref{sec:cs}.
For the weights used to balance the different terms in the GD cost function, see Tab.~\ref{tab:weights}.
In the self-similarity constraint (see Section~\ref{sec:selfsim}), we use a window size $\Lambda$ of $8$ (i.e.
half a bar), and for the tonality constraint we use an estimation window width $M$ of $4$ (see Section~\ref{sec:tonality}).
Fig.~\ref{fig:selfsim} shows some resulting samples, which are discussed in detail in Section \ref{sec:results-discussion}.

\subsubsection{Keyscape}\label{sec:keyscape}
We use keyscapes to illustrate the tonality of the pieces in Fig.~\ref{fig:iteration} and Fig.~\ref{fig:selfsim}.
A keyscape illustrates the tonal context over a musical piece, where each key receives a distinct colour.
We use the humdrum mkeyscape tool by David Huron, which analyses the musical piece with the Krumhansl-Schmuckler key-finding algorithm \cite{krumhansl1990} in different levels of detail.
The top of the pyramid depicts the key estimation for the entire piece, while towards the base the analysis is based on ever smaller window sizes.
Each scale has a distinct colour assigned to it and the keyscape is coloured according to the most predominant scale estimation.

\section{Results and discussion}\label{sec:results-discussion}
\subsection{Quantitative Evaluation}
Fig.~\ref{fig:info_rate} shows average Information Rates (IRs) for original Mozart piano sonatas and for samples from different models (cf. Section~\ref{sec:qu_evaluation}), where higher IRs indicate more distinct self-similarity structures.
It should be pointed out in advance that sampling from a probabilistic model introduces some sampling noise which increases the predictive entropy and therefore lowers the IR.
It is difficult to judge to what extent sampling noise on the one hand, and difficulties of the model to adapt to given constraints on the other hand, lead to the significantly lower IR of the constrained samples compared to the IR of the original Mozart transcriptions.
Nevertheless, it is clear from the results firstly that the IR of the original music is higher than that of the generated music, and secondly that the models \emph{without constraints} produce music with lower IR than the constrained C-RBM does.
Note that the latter point is a non-trivial result, since the self-similarity constraint does not explicitly optimise the Information Rate (neither do the tonal or meter constraints, obviously), but just encourages similarity or dissimilarity between the music at specific positions.
This result is in accordance with the initial observation that those models fail to generate higher-level self-similarity structure.
Due to their gating mechanism, GRUs are usually better at learning long-term dependencies than regular recurrent units.
The fact that the GRU-RBM does not perform better than the RNN-RBM shows that GRUs also have problems in modelling the content-invariant self-similarity property.

\subsection{Qualitative Evaluation}
Fig.~\ref{fig:selfsim} shows piano roll representations for the template piece (Fig.~\ref{fig:selfsim}[1a]), four generated samples that were constrained with properties from the template piece (Fig.~\ref{fig:selfsim}[2a] to Fig.~\ref{fig:selfsim}[5a]) and a baseline sample generated without constraints from the template piece (Fig.~\ref{fig:selfsim}[6a]).
The corresponding constraints for each musical piece are depicted in the respective figures b-d.
The repetition structure is marked on top of the template piece and over all other pieces and self-similarity matrices with vertical, green lines.\footnote{All samples illustrated in Fig.~\ref{fig:selfsim} can be listened to on Soundcloud under \url{http://www.soundcloud.com/pmgrbm}}

We chose the constrained samples by creating 20 solutions and picking the best four with respect to the minimal average value of the standardised GD cost and the standardised free energy over a constrained sampling process (see Section~\ref{sec:simulatedannealing} on standardising the cost and free energy functions).
Thus, results are selected to closely satisfy the given constraints rather than according to their musical quality.
Empirically we found that the musical quality in our setting increases when loosening the influence of the constraints, as this allows the probabilistic model to create more plausible samples (e.g.
the examples in Fig.~\ref{fig:selfsim} sometimes lack appropriate transitions between different sections which is an effect of both constraint satisfaction and limited training data).

By approximating the self-similarity matrix of the template piece, some aspects of the repetition structure were convincingly transferred to the constrained samples (see Fig.~\ref{fig:selfsim}[1b] to Fig.~\ref{fig:selfsim}[5b]).
For example, the exact repetitions C / C' and H / H' occur in every sample.
It is interesting to see how the extension of B to B' is solved.
Especially in the samples depicted in Fig.~\ref{fig:selfsim}[2] and Fig.~\ref{fig:selfsim}[3], the extension of B is realised by musical textures consistent with the immediate past.
In the sample in Fig.~\ref{fig:selfsim}[5], the model did not produce satisfactory results for phrase B and B' in a musical sense, although it found a solution which is self-similar over that time period and therefore satisfies that self-similarity constraint to a certain degree.

Parts E / E' / E'' are special cases, because even though they are very similar at first sight, they are transposed repetitions which cannot be captured by the self-similarity matrix as it is currently defined.
In the self-similarity matrix of the template, we see that each of those ``E'' sections is more or less similar or dissimilar to different regions in the piece.
In addition, we note that each repetition has the length of one bar.
When comparing these ``E'' sections with those in the samples, we recognise the limits of the method concerning temporal resolution.
The C-RBM has a filter length of one bar, which is too wide for sampling three bars with different requirements concerning similarity, while keeping a plausible low-level structure.
Therefore, in some samples the generated patterns span the whole, or at least two of the ``E'' sections.

Part G in the repetition structure is similar to most parts of the piece, as can be seen from the bright areas over the full height of the respective self-similarity matrices.
In the samples this is realised by choosing textures which are also similar to most parts.
Part J, in contrast, is very dissimilar to most areas of the template piece.
Probably due to limited training data, this sometimes results in empty areas in the samples.
Except for an apparent similarity in B and B', which is not reflected in the self-similarity matrix, the unconstrained baseline sample does not follow the repetition structure of the template piece.

The onset distributions (see Subplots \emph{c} in Fig.~\ref{fig:selfsim}), which are plots resulting from Equation~\ref{equ:onsetdist}, are sometimes rather dissimilar to the onset distribution of the template.
This shows that it is not easy to approximate this global property.
One reason for this may be that it is a property which summarises the complete music piece in only a few values, which makes it easy in the GD optimisation to approximate by distributing small changes over the whole sample.
Those are, however, locally not strong enough to be kept during GS.
Incorporating note durations for emphasizing onsets of longer notes would lead to more characteristic onset distributions, which could further lead to bigger local changes in the probability of notes in the piano roll.
However, in the onset distributions there is a tendency of the peaks at position $0$ and $8$ to be higher than the others, which corresponds to the tendency in the onset distribution of the template piece.
Note that the reason for every second value in the distributions being low is not the meter constraint but the stride of one beat in the convolution.
This provides the model with a regular grid allowing it to learn that the probability of an onset is lower at every second time step.
Therefore, those values are also low for the unconstrained baseline piece.

The keyscape (cf. Section~\ref{sec:keyscape}) for each sample is depicted in the respective subplots d.
We can see that the main key (i.e.
A major) of the template piece got transferred well to the constrained samples, as the colors of the upper areas of the keyscapes (purple) match exactly.
Towards the lower areas of the keyscapes, the colors of some samples do not correlate with those of the template piece's keyscape.
However, especially the modality to E major in the second quarter of the piece, depicted in red, and the blue area in the beginning of the piece (D major) are to some degree approximated in Fig.~\ref{fig:selfsim}[2d] and Fig.~\ref{fig:selfsim}[4d].
In sample Fig.~\ref{fig:selfsim}[3], the green area indicates a F\raisebox{.45mm}{$\sharp$} minor scale, which is similar to the E major scale (red) of the template piece (i.e.
there is a difference in one note, namely D/D\raisebox{.45mm}{$\sharp$}).
In general, the tonal structure of constrained samples is more stable than that of the unconstrained baseline sample, where the keyscape indicates tonal incoherence.

As mentioned above, the illustrated samples are the best four of $20$ with respect to the overall cost.
The most obvious shortcoming of samples not selected because of higher cost is that they do not satisfy some of the given structural constraints on a local level.
This includes the failure to reproduce a repetition at specific positions, or erroneously modulating into a key which does not occur in the template piece.
However, a closer inspection of such cases shows that incorrect keys are often closely related to the desired keys, for instance the parallel minor/major key.
Parallel minor/major keys have most of their pitches in common, but they are expected to follow a different distribution.
Another common problem in the non-optimal samples is that they show areas without any notes, which is probably a symptom of the contrasting objectives of GD and GS.
We found that the C-RBM is very sensitive to changes in the parameters of the whole system.
Other models able to perform GS could lead to a more stable functioning, like LSTMs used for GS in \cite{hadjeres2017deepbach}.


\section{Conclusion and future work}\label{sec:future}
Music is typically highly structured at both lower and higher levels.
State-of-the-art sequence models such as RNNs and LSTMs have been successfully used to generate music in restricted settings, but in more complex musical material, such as piano music from the classical or romantic period, not to mention orchestral works, important musical characteristics such as tonal, metrical and self-similarity structure tend to defy straight-forward time series modelling approaches.

The constrained sampling method for music generation presented here addresses this problem by combining a stochastic neural network for sampling plausible musical textures at a local level with soft constraints that impose higher-level structure regarding meter, tonality and self-similarity structure, obtained from a template piece.

The experimental validation of the proposed method reveals that the generated music possesses a stronger degree of structural organisation (as measured by Information Rate (IR)) than unconstrained models, including a state-of-the-art RNN-RBM model for polyphonic music generation.
A qualitative analysis of some generated music supports this finding, and clearly reveals repeated (but not identical) musical patterns, as well as global similarities in tonal structure to the template piece.

The empirical results also reveal some shortcomings.
Firstly, while imposing constraints with the proposed method helps to generate high-level structure, meaningful low-level structure can currently only be generated when the model is trained on relatively small amounts of data.
Overcoming this drawback may require more powerful generative models -- amenable to some form of Gibbs sampling -- as an alternative to the C-RBM.
Our hypothesis is that generative models can only generalise well on low-level structure if they are able to explicitly represent (transposed) repetition.
A promising approach to this is proposed by \citet{lattner2017relations}, who show that relations between musical sections can be learned and represented as so-called ``mapping codes''.

Secondly, when listening to the generated musical samples it is clear that the tonal, meter and self-similarity constraints presented here are by no means fully elaborated nor exhaustive.
For example, more specialised constraints -- like a differentiable formalisation of the IR measure -- could optimise sequences to directly obey desired structural properties.
Perhaps most importantly, what is currently missing is a constraint that enforces musical closure at boundaries of structural units.
Without such a constraint, the music contains repeated musical structures, but these structures are hard to identify perceptually because their boundaries are not marked by salient musical cues (such as harmonic resolution).
That said, the proposed constrained sampling approach is general enough to accommodate this and possibly other constraints.

\subsubsection*{Acknowledgements}
This research was supported by the EU FP7 (project Lrn2Cre8, FET grant number 610859), and the European Research Council (project CON ESPRESSIONE, ERC grant number 670035).


\printbibliography



\end{document}